\documentclass[mathleft]{an}
\usepackage{graphicx}
\usepackage{times}
\begin{document}
\Pagespan{789}{}
\Yearpublication{2007}%
\Yearsubmission{2007}%
\Month{11}%
\Volume{999}%
\Issue{88}%

\title{Anti-solar differential rotation and surface flow pattern on UZ Libr\ae}

\author{K.~Vida\inst{1,2}\fnmsep\thanks{Corresponding author:
  \email{k.vida@elte.hu}\newline}
\and Zs.~K\H{o}v\'ari\inst{2}
\and M.~\v{S}vanda\inst{3,4}
\and K.~Ol\'ah\inst{2}
\and K.~G.~Strassmeier\inst{5}
\and J.~Bartus\inst{5}
\and E.~Forg\'acs-Dajka\inst{1}
}

\institute{
E\"otv\"os University,
Department of Astronomy, H-1518 Budapest, P.O.Box. 32, Hungary
\and
Konkoly Observatory, H-1525 Budapest, P.O.Box 67, Hungary
\and Astronomical Institute, Academy of Sciences of the Czech
Republic, v.~v.~i., Fri\v{c}ova 298, 251 65 Ond\v{r}ejov, Czech
Republic
\and Astronomical Institute, Faculty of Mathematics and
Physics, Charles University in Prague, V Hole\v{s}ovi\v{c}k\'{a}ch
2, 180 00 Prague~8, Czech Republic
\and Astrophysical Institute Potsdam, An der Sternwarte
16, D-14482 Potsdam, Germany 
}

\received{xx xxx 2007}
\accepted{xx xxx 2007}
\publonline{later}


\abstract{We re-investigate UZ Libr\ae~spectra obtained at KPNO in 1998 and 2000.
From the 1998 data we compose 11 consecutive Doppler images using the Ca\,{\sc i}-6439,
Fe\,{\sc i}-6393  and Fe\,{\sc i}-6411 lines. Applying the method of average cross-correlation
of contiguous Doppler images we find
anti-solar differential rotation with a surface shear of $\alpha\approx-0.03$. The pilot application
of the local correlation tracking technique for the same data qualitatively
confirms this result and indicates complex flow pattern on the stellar surface.
From the cross-correlation of the two available Doppler images in 2000 we also
get anti-solar differential rotation but with a much weaker shear of $\alpha\approx-0.004$.}
\maketitle

\sloppy
\section{Introduction}
In our time, the Sun is the only star for which
the surface differential rotation and other surface flows can be measured directly.
The detection of differential rotation on active stars, however, is a
challenging observational task. Starspots are regarded
as manifestations of solar-like active regions
and as such, are suitable for tracking and
recovering surface flows. In this paper we make one of the first attempts at recovering
surface flows on spotted stars (see also
the paper by K\H{o}v\'ari et al. \cite{sgemposter}).

Our target is the RS\,CVn-type active giant star UZ\,Libr\ae.
The star has a very fast rotation rate of $P_\mathrm{rot}=4.76$ days, which is equivalent to the orbital period.
%
Doppler images of the star were presented by Strassmeier (\cite{1996A&A...314..558S}) and
Ol\'ah, Strassmeier \& Weber (\cite{2002A&A...389..202O}). In the latter paper the signature of a weak anti-solar differential rotation (DR) was found but the authors regarded the detection inconclusive.
The possible DR was also examined by Ol\'ah Jurcsik and Strassmeier (\cite{2003A&A...410..685O}) using photometric observations. They found two permanent, distinct periods in the Fourier-spectrum of the long-term data,
which were assigned to the different rotation rate of spots at high
and low latitudes, resulting in $\alpha=\Delta\Omega/\Omega$ of $-0.0026$. Adopted stellar parameters of UZ\,Lib
are summarised in Table \ref{tab:params}, the values which are not marked were taken from Ol\'ah, Strassmeier \& Weber (\cite{2002A&A...389..202O}).


\begin{table}[]
\caption{Stellar parameters of UZ Libr\ae}\label{tab:params}
\leavevmode
\footnotesize
\begin{tabular}{ll}
\noalign{\smallskip}
\hline
\noalign{\smallskip}
  spectral type$~^1$ &K0 III \\
  $P_{rot}=P_{orb}~^{2,3}$ & 4.768241 days \\
  $v\sin i~^{2,3}$ & 67 $\pm$ 1 km\,s$^{-1}$\\
  $i~^4$ & $50^\circ\pm$10\degr\\
  T$_\mathrm{eff}~^3$ & 4800 K\\
  $\log g~^2$ & 2.5\\
  Microturbulence $\xi$& 0.8 km s$^{-1}$\\
  Macroturbulence $\zeta_\mathrm{R}=\zeta_\mathrm{T} $ & 4 km s$^{-1}$\\
  Chemical abundance & solar\\
  \noalign{\smallskip}
  \hline\\[-5pt]
\end{tabular}

\noindent$^1$ Bopp \& Stencel (\cite{1981ApJ...247L.131B})\\
\noindent$^2$ Strassmeier (\cite{1996A&A...314..558S})\\
\noindent$^3$ Fekel et al. (\cite{1999A&AS..137..369F})
\end{table}

\section{Data processing}
We used two sets of spectra obtained at Kitt Peak National Observatory (KPNO) in 1998 and
2000. 
From the 16 spectra obtained in 1998 we recover a time-series of 11 Doppler images.
For the first image we take the first six spectra, and run our inversion code
\textsc{TempMap} (Rice et al. \cite{1989A&A...208..179R}) for the Ca\,{\sc i}-6439,
Fe\,{\sc i}-6393 and Fe\,{\sc i}-6411 mapping lines.
Then, we proceed with the next set of six spectra, leaving the first spectrum and
adding the subsequent one (i.e, the 7th) to the modelled datased, producing a slightly different
Doppler image. This pattern is followed until we get the last image from the last
six spectra of the time-series. The respective Doppler images for the three different
mapping lines are in quite good agreement with each other and are averaged for
further use, resulting in a time-series of 11 Doppler images covering about 20 days
(4.2 rotation cycles). Note, that a similar method was applied to get a motion picture of the
surface spots on HR\,1099 (Strassmeier \& Bartus \cite{2000A&A...354..537S}) and
later on LQ\,Hya (K\H{o}v\'ari et al. \cite{2004A&A...417.1047K}).
From the 17 spectra gathered in 2000 we reconstructed only two images from two subsets
consisting of 8 and 9 spectra, both covering about 8 days, but separated by a time lag
of 16 days (Ol\'ah, Strassmeier \& Weber \cite{2002A&A...389..202O}).

\section{Results from the cross-correlation study}

The method of `Average Cross-CORrelation of contiguous Doppler images' (hereafter ACCORD)
uses a combination of independent image pairs (i.e., which do not have common spectra),
whose relevant latitude strips are cross-correlated with each other. When having multiple
cross-correlation maps, the elapsed time lags between the correlated
images are different, so in order to average them, one has to normalize the cross-correlation
maps (i.e., applying a linear stretching or compression along the longitude shift) depending on 
the time lag. Unfortunately, the larger the time lag between the
cross-correlated images, the greater the risks of masking the DR pattern by individual
spot evolutions (waxing or waning of spots, emerging of new spots, melting, etc.). Thus,
to minimize this effect, we correlate only the consecutive but contiguous (i.e., the
closest independent) images. The average cross-correlation map is then fitted by a
quadratic differential rotation law:
$\Omega (\vartheta)=\Omega_\mathrm{eq}-\Delta\Omega\sin^2\vartheta$,
where $\vartheta$ stands for latitude, $\Omega_\mathrm{eq}$ means the angular velocity
on the equator, and $\Delta\Omega=\Omega_\mathrm{eq}-\Omega_\mathrm{pole}$ the
angular velocity difference between the equator and the poles. The differential rotation
parameter is derived as $\alpha=\Delta\Omega/\Omega_{eq}.$
For a detailed description of this method, as well as for further applications
see K\H{o}v\'ari et al. (\cite{2004A&A...417.1047K}, \cite{AARNsgem}).

\begin{figure}[t]
 \includegraphics[angle=0,width=0.48\textwidth]{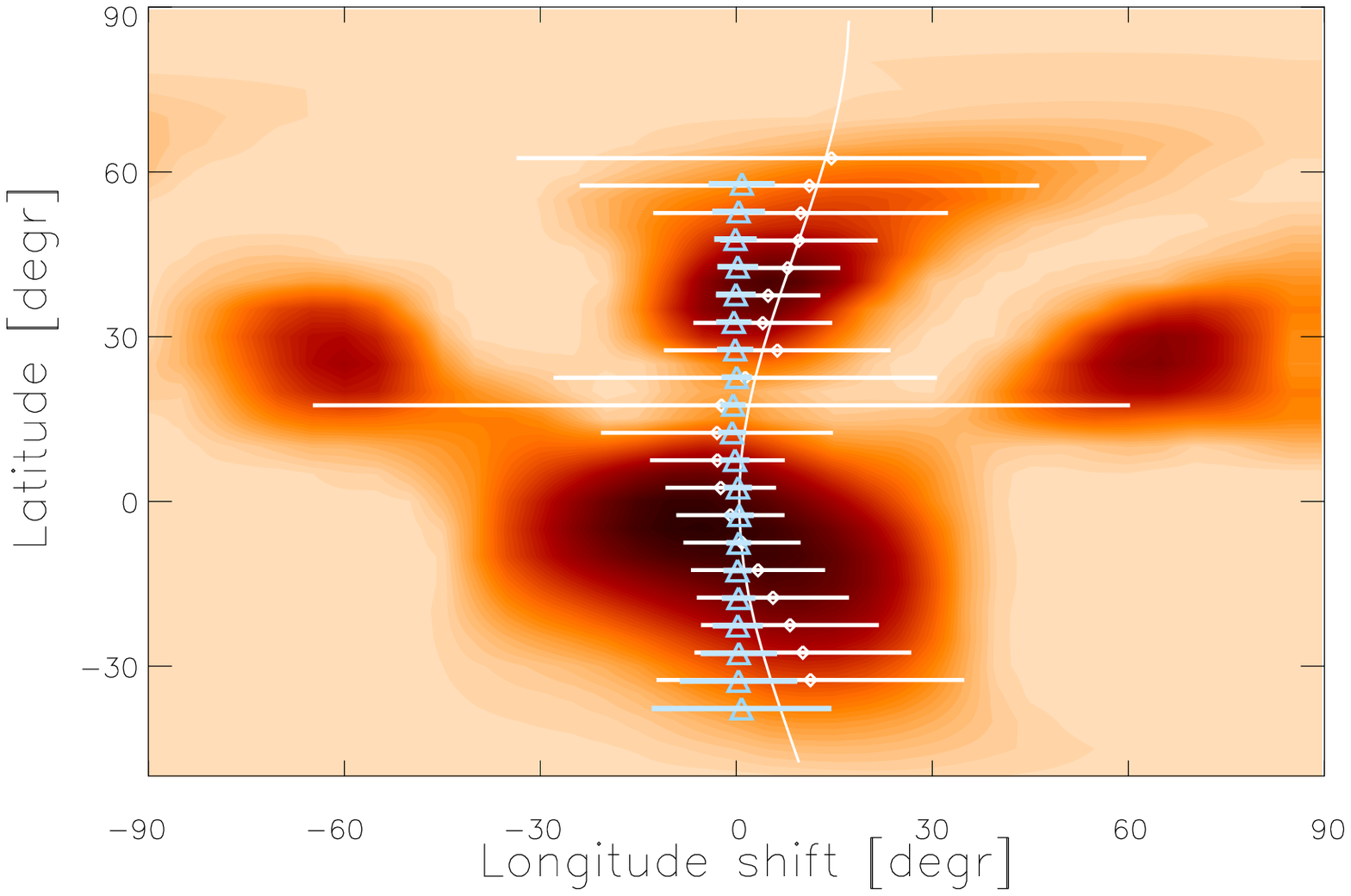}

 \includegraphics[angle=0,width=0.48\textwidth]{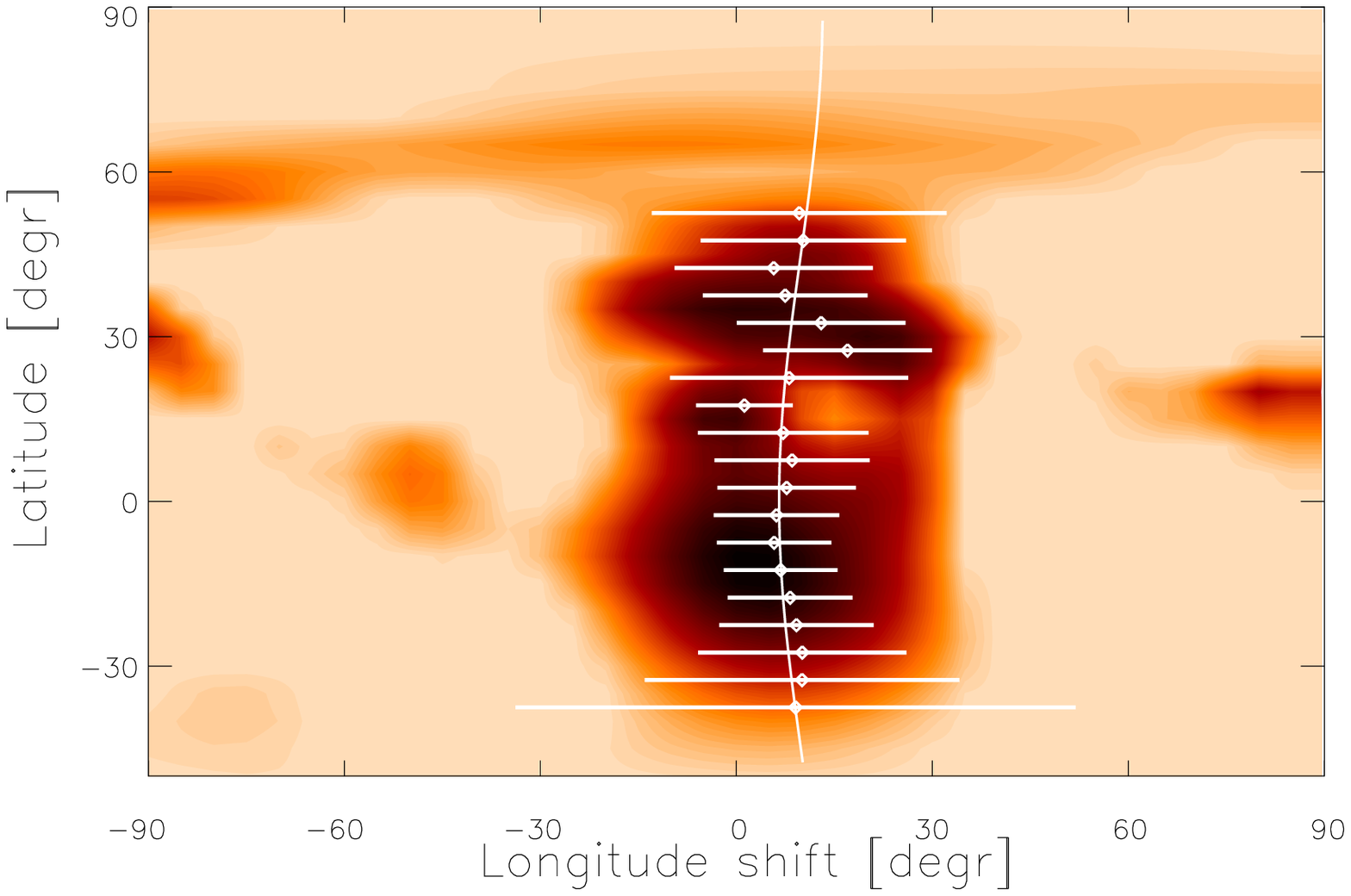}

\caption{Cross-correlation maps obtained from 11 time-series Doppler images in 1998 (top)
and from 2 Doppler images in 2000 (bottom). The better the correlation the
darker the shade. Cross-correlation functions per latitude are fitted by Gaussian curves
(Gaussian peaks are indicated by dots, the FWHMs by error bars).
The continuous line shows the fitted quadratic differential rotation law.
Longitude shifts from the LCT method are overplotted as
triangles in the top panel (see Sect.~\ref{sect:lct} for details).}
\label{fig:ccfits}
\end{figure}

\begin{figure*}
\includegraphics[angle=0,width=0.88\textwidth]{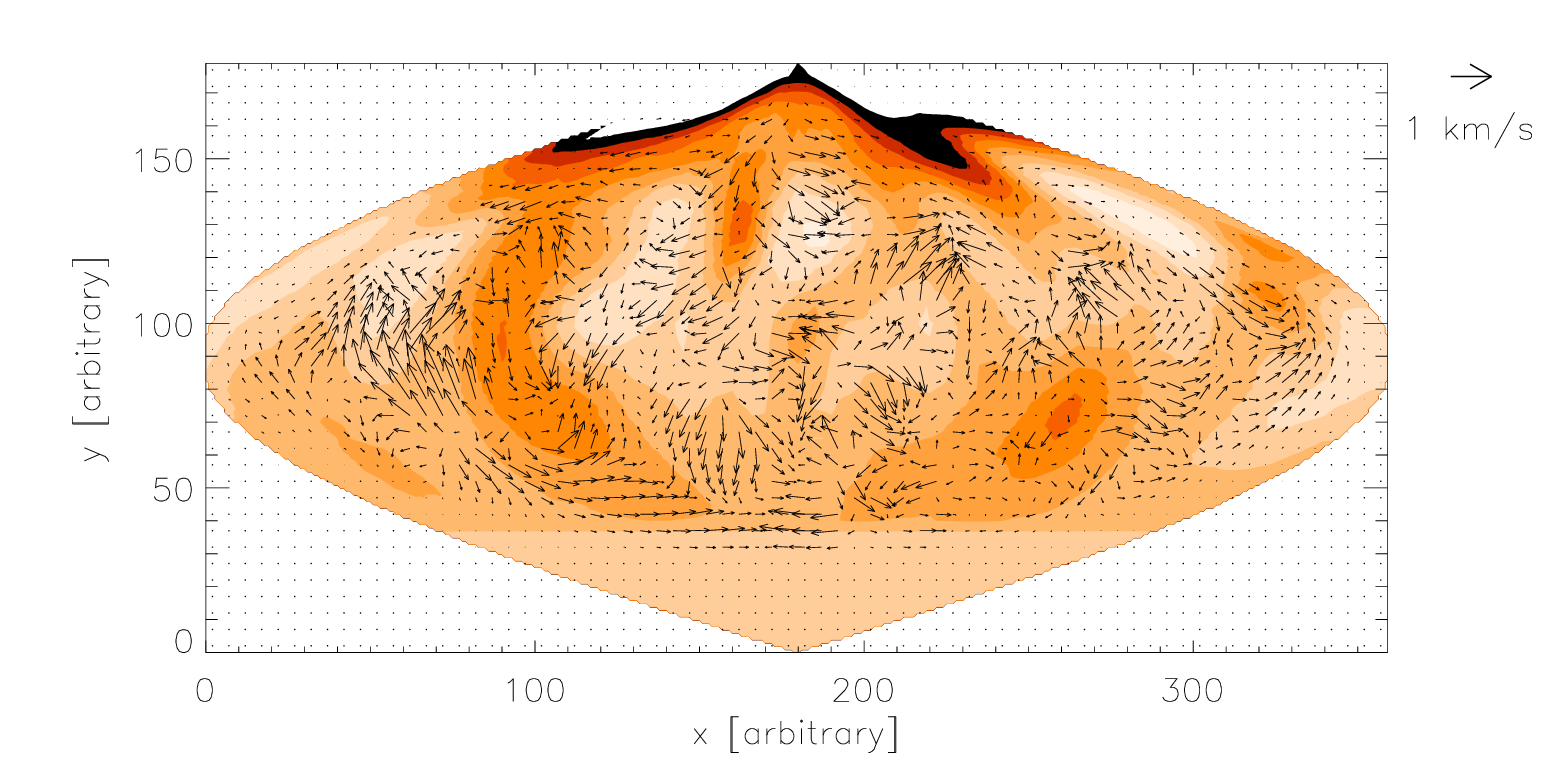}\hspace{-0.95cm}
\includegraphics[angle=0,height=7.4cm]{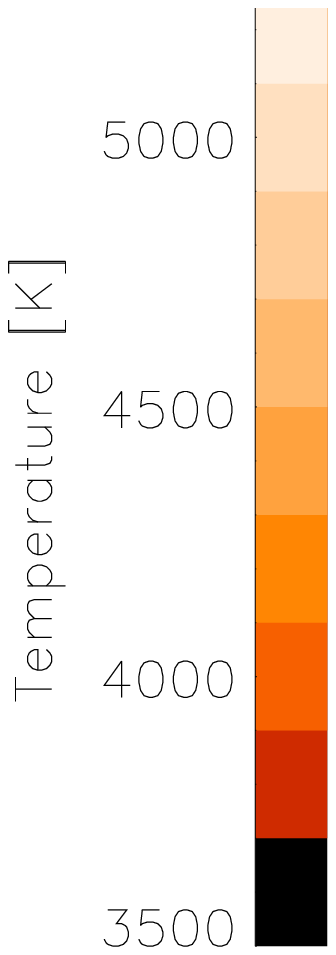}
\caption{The result of the LCT technique showing the surface flow pattern on UZ Libr\ae\ 
from the time-series Doppler images in 1998. As a background the average surface
temperature map is plotted.}
\label{fig:sv_flow}
\end{figure*}

The resulting cross-correlation maps are plotted in Fig.~\ref{fig:ccfits} for 1998 (top) and for 2000 (bottom).
The best correlated longitude shifts are fitted by a quadratic DR law.
The best-fit rotation function indicates zero longitude shift of the equatorial zone, i.e., the equator is
rotating with the orbital period. This result agrees well with the findings in
Ol\'ah, \hbox{Jurcsik} \& Strassmeier (\cite{2003A&A...410..685O}). The derived $\alpha$
parameters are $-0.027\pm0.003$ and $-0.004\pm0.001$ for the 1998 and 2000 data, respectively,
indicating a weak anti-solar differential rotation in both cases, matching (at least qualitatively) the negative DR parameter derived from photometric
data by Ol\'ah, Jurcsik \& Strassmeier (\cite{2003A&A...410..685O}).

The cross-correlation study can be repeated  for the corresponding longitudinal
stripes (meridian circles). This case we take only the hemisphere of the always
visible pole, because Doppler imaging is more reliable for this part. We then look
for the correlation pattern in the same way when searching for DR from longitudinal cross-correlation.
The resulting latitudinal cross-correlation pattern is shown in Fig.~\ref{fig:merflows} (top).
The best correlated latitude shifts indicate incoherent flows, with too large error bars.

\section{Results from local correlation tracking}\label{sect:lct}

We also apply the local correlation tracking (hereafter LCT) technique, which was originally designed
to remove seeing-induced distortion in sequences of solar images. This technique examines the best
match between the pixels of the consecutive images in a correlation window. The best match defines
a vector of displacement, which, as a result gives a map of surface flow patterns.
For further details of the method as well as for application see
\v{S}vanda et al. (\cite{2006A&A...458..301S}, \cite{2007SoPh..241...27S}) and 
our forthcoming paper for numerical tests on artificial data
by \v{S}vanda et al. (\cite{lctlq}).

The resulting surface flow map for the 1998 data is plotted in Fig.~\ref{fig:sv_flow}, indicating a complex
flow pattern with convergent and also divergent flows, and maybe vortex-like features around the spots near $100^\circ$.
The differential rotation pattern can be derived from averaging the
zonal flow components. The best-fit numerical result suggests a weak but uncertain anti-solar shear of $-0.001\pm0.003$ which is at least not in a contradiction with the ACCORD results. The latitude shift -- longitude diagram from LCT
is overplotted on the corresponding ccf-map in Fig.~\ref{fig:ccfits} with triangles.
Unfortunately, the LCT technique cannot be applied for the two separated subsets in 2000.

\begin{figure}
 \includegraphics[angle=-0,width=1.00\columnwidth,height=0.3\columnwidth]{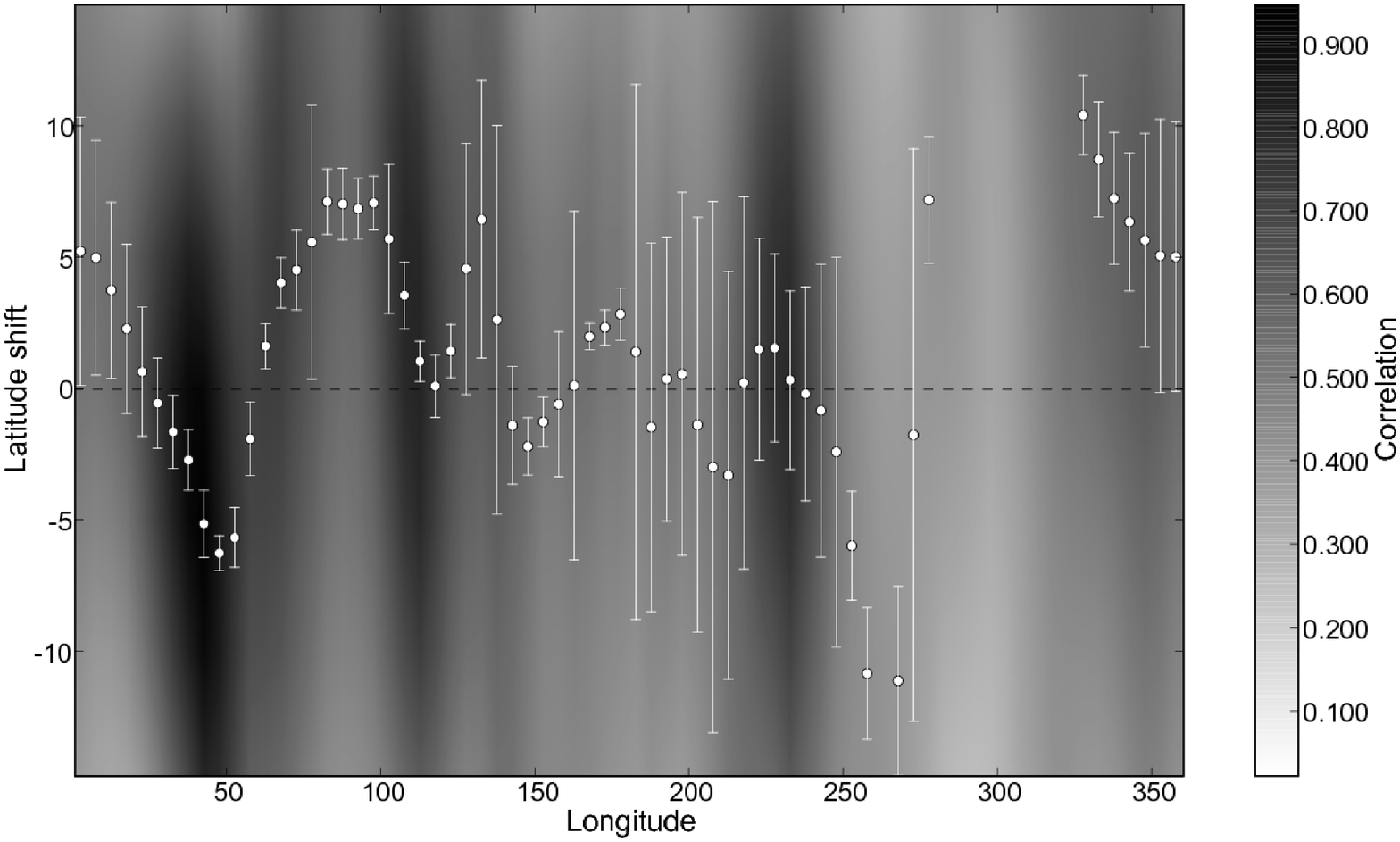}

\hspace*{-0.22cm} \includegraphics[angle=-0,width=0.88\columnwidth,height=0.3\columnwidth]{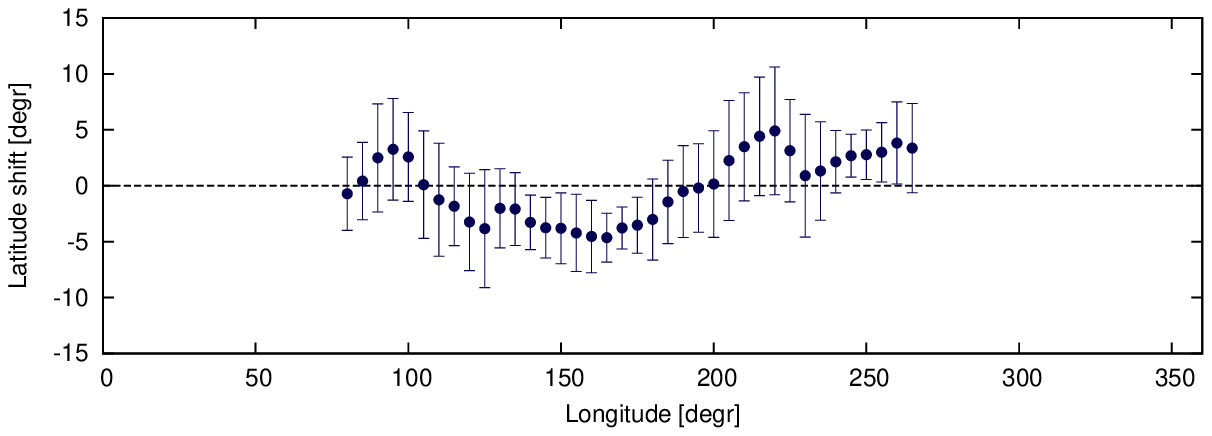}
\caption{Best correlated latitudinal shifts from ACCORD (top) and meridional flow components from
LCT (bottom). No sign of coherent meridional flow can be seen.}
\label{fig:merflows}
\end{figure}

From averaging the meridional flow components we are able to investigate the
meridional motions. The resulting plot is seen in Fig.~\ref{fig:merflows} (bottom).
We get small velocity values in the order of a few hundreds m\,s$^{-1}$, and
the comparison with the ACCORD result is meaningless.

\section{Summary}

\begin{itemize}
\item Applying our cross-correlation technique ACCORD for the time-series Dopper images of UZ\,Lib
in 1998 and in 2000 we derived weak surface DR of anti-solar type,
in agreement with former DR detections. However, no
sign of coherent meridional flow was found.
\item LCT technique is applied first time for stellar data (see also
the paper by K\H{o}v\'ari et al. \cite{sgemposter}) and delineated
a complex surface flow network for UZ\,Lib.
\item Averaging the zonal components of the LCT flow map along latitude 
resulted in an uncertain, weak anti-solar DR.
Meridional flow components show no evidence for any coherent pattern, similarly to the ACCORD non-detection.

\end{itemize}

\acknowledgements
KV appreciates the hospitality of Konkoly Observatory.
KV, ZsK, and KO are grateful to the Hungarian Science
Research Program (OTKA) for support under the grant T-048961.
ZsK  is a grantee of the Bolyai J\'anos Scholarship of the Hungarian
Academy of Sciences. M\v{S} is grateful for support by Research Program MSM0021620860 of the Ministry
of Education of the Czech Republic.


\begin{thebibliography}{}

\bibitem[1981]{1981ApJ...247L.131B} Bopp, B.~W., 
Stencel, R.~E.:\ 1981, \apjl 247, L131 

\bibitem[1999]{1999A&AS..137..369F} Fekel, F.~C., 
Strassmeier, K.~G., Weber, M., Washuettl, A.:\ 1999, \aas 137, 369 

\bibitem[2004a]{2004PADEU..14..221K} 
K\H{o}v\'ari, Zs., Weber, M.:\ 2004, PADEU Vol 14, 221 

\bibitem[2004]{2004A&A...417.1047K} K\H{o}v\'ari, Zs., 
Strassmeier, K.~G., Granzer, T., Weber, M., Ol{\'a}h, K., Rice, J.~B.:\ 
2004, \aaa 417, 1047

\bibitem[2007a]{sgemposter}
K\H{o}v\'ari, Zs., Bartus, J., Strassmeier, K.~G., Vida, K., \v{S}vanda, M.,
Ol\'ah, K.: 2007a, AN, this volume

\bibitem[2007b]{AARNsgem}
K\H{o}v\'ari, Zs., Bartus, J., Strassmeier, K.~G., Vida, K., \v{S}vanda, M.,
Ol\'ah, K.: 2007b, A\&A 474, 165

\bibitem[2002]{2002A&A...389..202O} Ol{\'a}h, K., 
Strassmeier, K.~G., Weber, M.:\ 2002, \aaa 389, 202 


\bibitem[2003]{2003A&A...410..685O} Ol{\'a}h, K., Jurcsik, 
J., Strassmeier, K.~G.:\ 2003, \aaa 410, 685

\bibitem[1989]{1989A&A...208..179R} Rice, J.~B., Wehlau, 
W.~H., Khokhlova, V.~L.:\ 1989, \aaa 208, 179 

\bibitem[1996]{1996A&A...314..558S} Strassmeier, K.~G.:\ 1996, 
\aaa 314, 558 

\bibitem[2000]{2000A&A...354..537S} Strassmeier, 
K.~G., Bartus, J.:\ 2000, \aaa 354, 537 

\bibitem[2006]{2006A&A...458..301S} {\v S}vanda, M., 
Klva{\v n}a, M., Sobotka, M.:\ 2006, \aaa 458, 301 

\bibitem[2007]{2007SoPh..241...27S} {\v S}vanda, M., 
Zhao, J., Kosovichev, A.~G.:\ 2007, Sol. Phys. 241, 27 

\bibitem[2008]{lctlq}
\v{S}vanda, M., K\H{o}v\'ari, Zs., Strassmeier, K.~G.: 2008, A\&A,
in prep.

\end{thebibliography}
\end{document}